\documentclass{appolb}
\usepackage{epsfig}
\begin{document}
\pagestyle{plain}
\newcount\eLiNe\eLiNe=\inputlineno\advance\eLiNe by -1
\title{Assortativity in random line graphs
\thanks{Presented on the Summer Solstice 2009 International Conference on Discrete Models of Complex Systems, Gda\'nsk, Poland, 22-24 June 2009. Correspondence to: {\tt kulakowski@novell.ftj.agh.edu.pl}}}
\author{ Anna Ma\'nka-Kraso\'n and Krzysztof Ku{\l}akowski
\address{Faculty of Physics and Applied Computer Science, AGH University of
 Science and Technology, al. Mickiewicza 30, PL-30059 Krak\'ow,
 Poland
}}

\maketitle
\begin{abstract}
We investigate the degree-degree correlations in the Erd\"os-R\'enyi networks, the growing exponential networks and the scale-free networks. 
We demonstrate that these correlations are the largest for the exponential networks. We calculate also these correlations 
in the line graphs, formed from the considered networks. Theoretical and numerical results indicate that all the line graphs
are assortative, i.e. the degree-degree correlation is positive.
\end{abstract}
\PACS{64.60.aq; 02.10.Ox; 05.10.Ln}
  
\section{Introduction}
A network of tennis players is formed when we link two players who met in the same game. Alternatively we 
can form a network of tennis games; two games are linked if the same competitor played in both of them. The same can be told 
on boxers and football teams. This construction is known as a line graph \cite{har,bal,wiki}. Each graph can be converted to 
its line graph. Under this transformation links become nodes, and two nodes
of the line graph are linked if the respective links in the original graph share a node. The mathematical representation 
of a network by its line graph can be of interest in the science of complex networks \cite{b0}; for some applications of 
line graphs see \cite{mokre,lij,evla,nach1,nach2,chi,chi2}. \\

Our concern in line graphs is due to their specific topology. Recently we shown that line graphs formed from the 
Erd\"os-R\'enyi networks, the growing exponential networks and the Barab\'asi-Albert scale-free networks are highly 
clustered, with the clustering coefficient $C$ higher than 0.5 \cite{my}. This makes the line graphs to be potentially
attractive for modeling of social networks, which are also highly clustered \cite{newrec}. Here we focus on the 
degree-degree correlation in the line graphs, formed from the three kinds of networks listed above. Once this correlation
is positive, nodes of high degree are more frequently linked to nodes of high degree; such networks are termed to show
assortative mixing \cite{new2}. If the degree-degree correlation is negative, the mixing is termed disassortative.\\

 \begin{figure}[ht]
 %\vspace{0.3cm}
 \centering
 {\centering \resizebox*{12cm}{9cm}{\rotatebox{-90}{\includegraphics{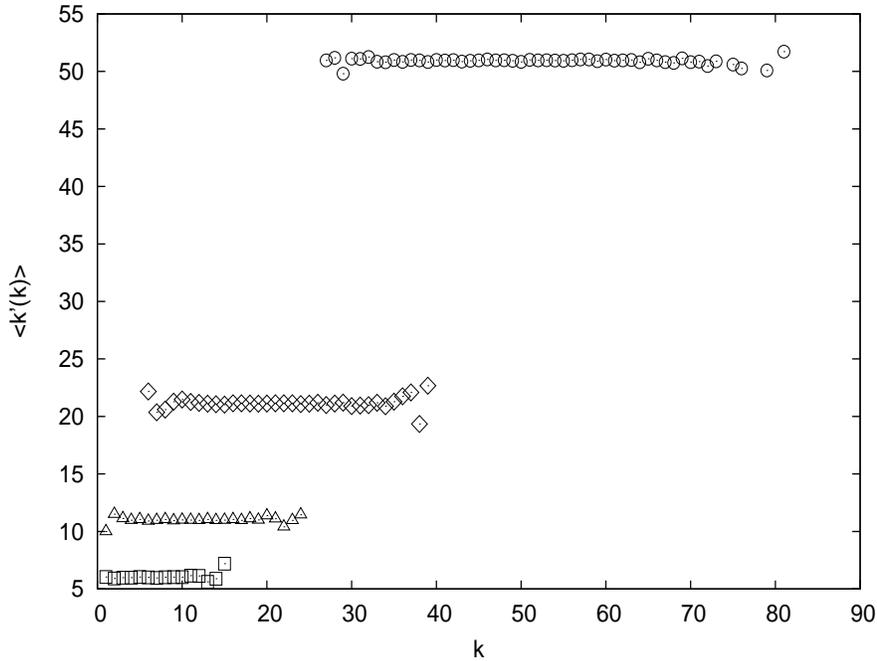}}}}
 %\vspace{0.3cm}
 \caption{Degree-degree correlations in the Erd\"os-R\'enyi networks, measured by the curves $\left\langle k'(k)\right\rangle$, where $k'$ is the degree of a neighbor of a node of degree $k$. The data shown are obtained for $\left\langle k\right\rangle=5,10,20$ and $50$ (squares, triangles, rhombs and circles, respectively). The data increase with $\left\langle k\right\rangle$.}
 \label{fig-1}
 \end{figure}

In our former calculations \cite{my}, theoretical calculations of the clustering coefficient $C$ in the line graphs
were based on the assumption, that there is no degree-degree correlations in the initial networks. The accordance 
of the theoretical results with the simulations was quite reasonable at least for well connected networks, with mean
degree larger than 10. Still, some differences could be observed for the exponential networks (Fig. 4 in \cite{my}).
Our aim here is $\it i)$ to compare the degree-degree correlations in the the Erd\"os-R\'enyi networks, the exponential 
networks and the scale-free networks, $\it ii)$ to calculate these correlations in the line graphs, obtained from the 
three kinds of networks. The results are presented in the form of $\left\langle k'(k)\right\rangle$, where $k'(k)$ is the degree of a neighbour
of a node of degree $k$.\\

Next section is devoted to the numerical calculations of the degree-degree correlations in the initial networks. In section 3, analytical 
calculations of $\left\langle k'(k)\right\rangle$ for the line graphs are presented. In Section 4 we show the correlations in the line graphs, obtained  numerically. Last section is devoted to conclusions.

\section{Numerical calculations for the initial networks}

The original Erd\"os-R\'enyi network is generated from $N=10^4$ nodes; a link is placed between two nodes with the probability $p$.
For the exponential and scale-free networks the algorithm starts from a fully connected cluster of $M$ nodes. In a series of steps new nodes are added, each with $M$. Each edge of this node is connected to a randomly chosen node. For the scale-free network we have to use preferential attachment; nodes are selected proportionally to their degree. The final size of the exponential and scale-free networks is again $N=10^4$ nodes.\\

 \begin{figure}[ht]
 %\vspace{0.3cm}
 \centering
 {\centering \resizebox*{12cm}{9cm}{\rotatebox{-90}{\includegraphics{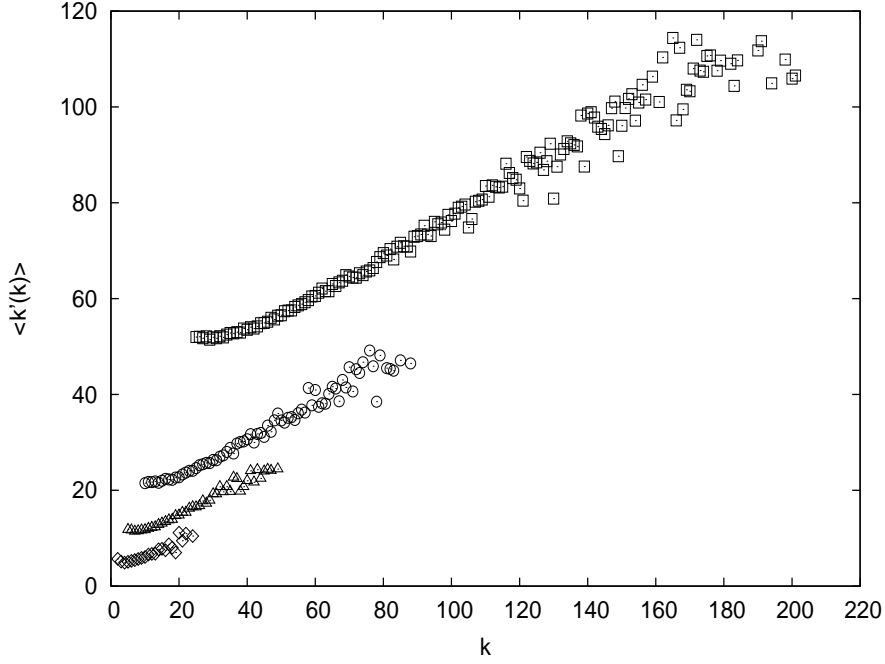}}}}
 %\vspace{0.3cm}
 \caption{Degree-degree correlations in the growing exponential networks, measured by the curves $\left\langle k'(k)\right\rangle$, where $k'$ is the degree of a neighbor of a node of degree $k$. The data shown are obtained for $\left\langle k\right\rangle=4,10,20$ and $50$ (rhombs, triangles, circles and squares, respectively). The data increase with $\left\langle k\right\rangle$.}
 \label{fig-1}
 \end{figure}

To evaluate the degree-degree correlation we check how the average degree $k'$ of the nearest neighbours of nodes with degree $k$ depends on $k$. Numerical calculation begins with a search for nodes with degree $k$. Then, the average degree is calculated of all nearest neighbours of these nodes.
Those steps are repeated for subsequent values of $k$.\\

 \begin{figure}[ht]
 %\vspace{0.3cm}
 \centering
 {\centering \resizebox*{12cm}{9cm}{\rotatebox{-90}{\includegraphics{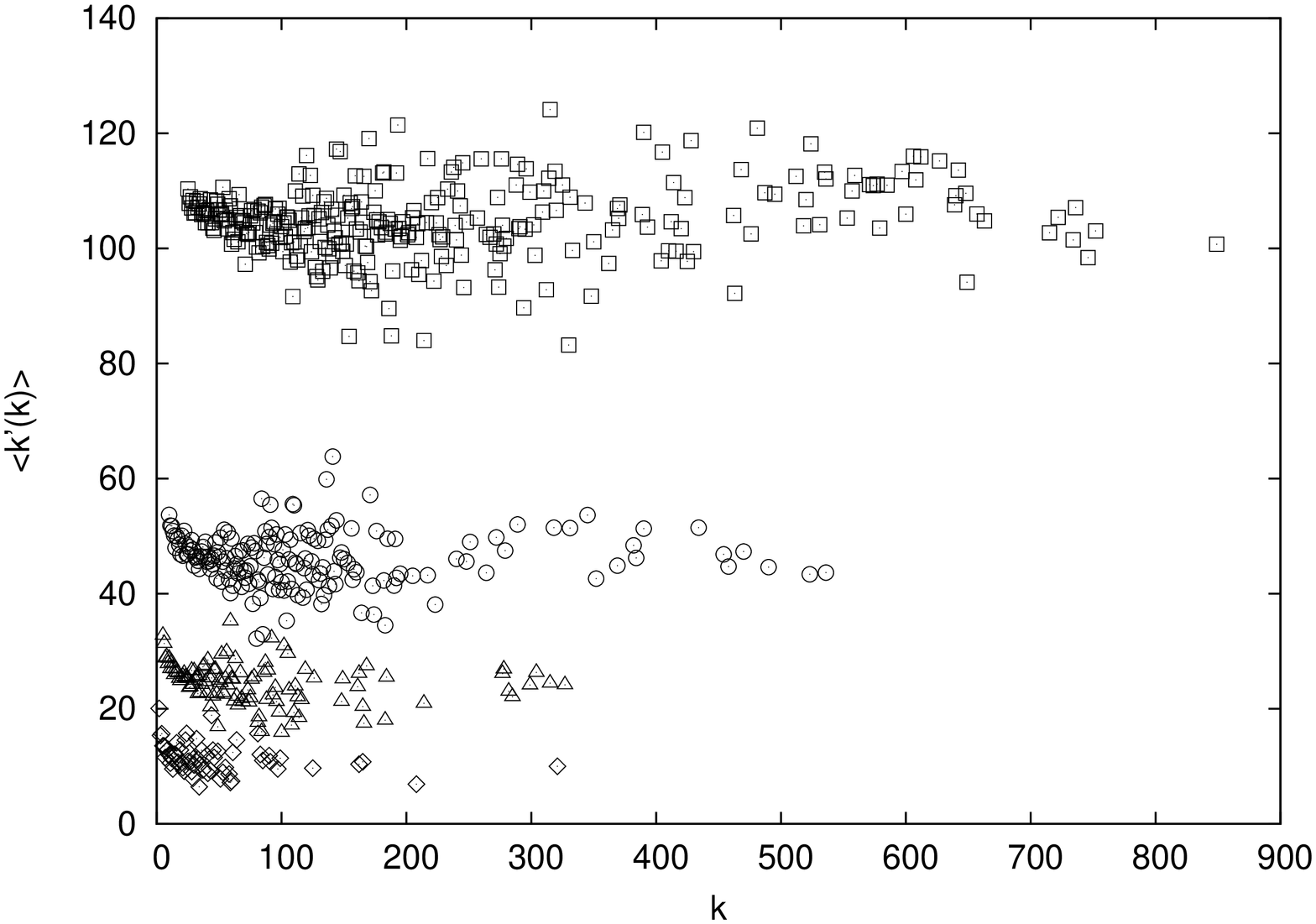}}}}
 %\vspace{0.3cm}
 \caption{Degree-degree correlations in the growing Barab\'asi-Albert networks, measured by the curves $<k'(k)>$, where $k'$ is the degree of a neighbor of a node of degree $k$. The data shown are obtained for $\left\langle k\right\rangle=4,10,20$ and $50$ (rhombs, triangles, circles and squares, respectively). The data increase with $\left\langle k\right\rangle$.}
 \label{fig-1}
 \end{figure}

As it is shown in Fig. 1, the slope of obtained curves is close to zero. This means, that the Erd\"os-R\'enyi networks show no degree-degree correlations, i.e. no assortativity at all. This result is a natural consequance of the construction of this kind of networks. On the contrary, the results for the exponential networks (Fig. 2) indicate that the degree-degree correlations are positive: more connected nodes are nearest neighbours of also more connected ones. The result is in accordance with analytical calculations \cite{new,qi}. The results for the Barab\'asi-Albert networks (Fig. 3) are more fuzzy. Still, except perhaps the case of small $k$'s, the correlations are not observed. This observation coincides with the conclusion of \cite{new2}, obtained from analytical method. 

 \section{Analytical calculations for the line graphs}

The assortativity of the line graph is to be investigated by the calculation of the mean degree of a node, converted 
from a link, which is a neighbour of another node of degree $k$, converted also from a link. These two links shared 
a node in the initial graph. The notation is as follows: the first link joined nodes of degrees $k_1$ and $k_2$, and 
the second link joined nodes of degrees $k_2$ and $k_3$. Now these links are nodes, with degrees $k_1+k_2-2$ and $k_2+k_3-2$,
respectively. We assume that there is no degree-degree correlations in the initial graph. Then the mean degree $\left\langle k'(k)\right\rangle$ of a neighbour
of a node of degree $k$ in the line graph can be found as

\begin{equation}
\left\langle k'(k)\right\rangle=\frac{\sum_{k_1,k_2,k_3}k_1P(k_1)k_2P(k_2)k_3P(k_3)(k_1+k_2-2)\delta_{k,k_2+k_3-2}}{\sum_{k_1,k_2,k_3}k_1P(k_1)k_2P(k_2)k_3P(k_3)\delta_{k,k_2+k_3-2}}
\end{equation}
where $P(k)$ is the degree distribution for the initial graph. We use the Kronecker delta to eliminate the sums over $k_2$. Then, the sums
over $k_1$ are from one to infinity, and the sums over $k_3$ from one to $k$.\\

 \begin{figure}[ht]
 %\vspace{0.3cm}
 \centering
 {\centering \resizebox*{12cm}{9cm}{\rotatebox{-90}{\includegraphics{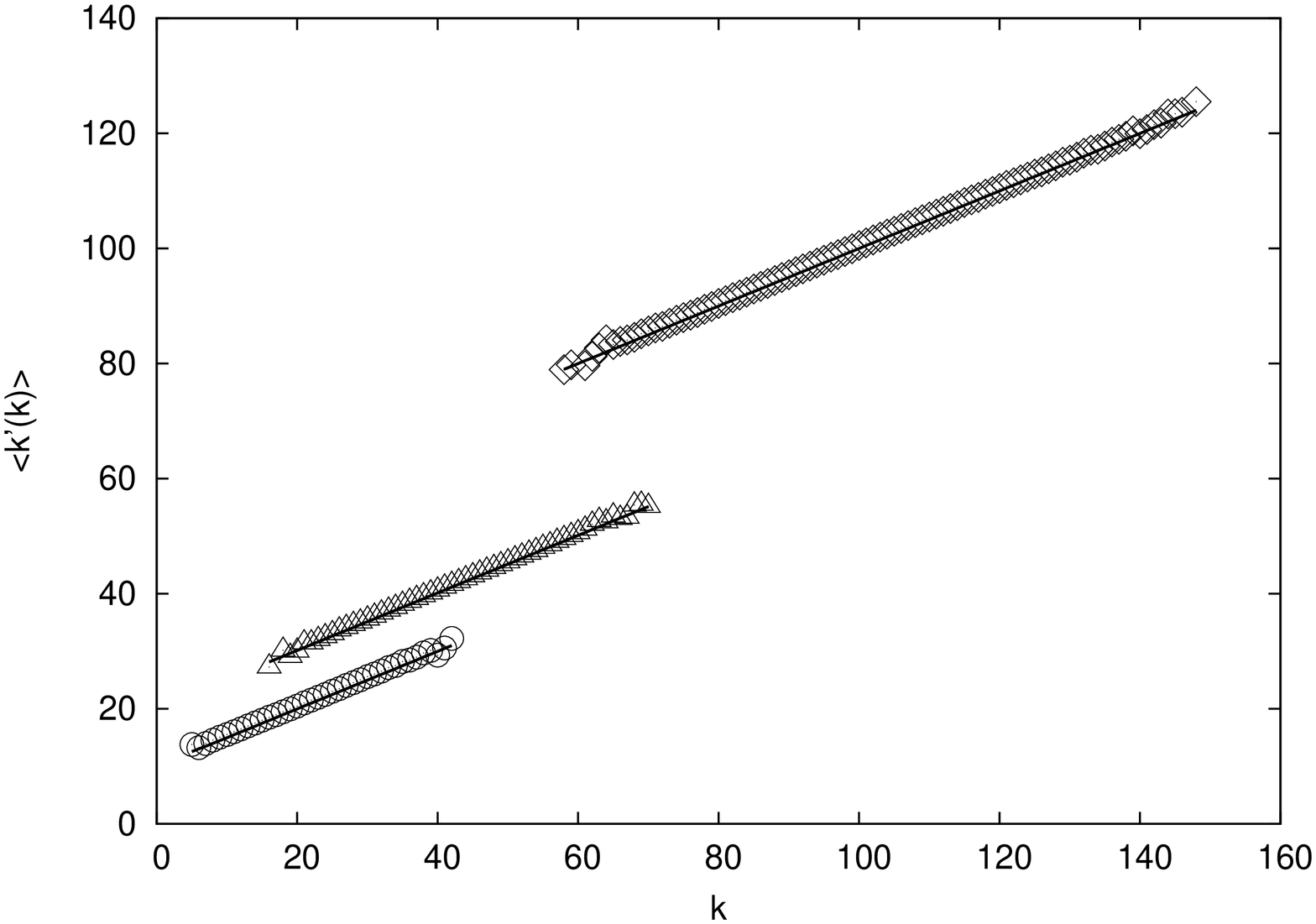}}}}
 %\vspace{0.3cm}
 \caption{Degree-degree correlations in the line graphs constructed from the Erd\"os-R\'enyi networks. The data shown are obtained for $\left\langle k\right\rangle=10,20$ and $50$ (circles, triangles and rhombs, respectively). Lines are obtained from Eq. 2.}
 \label{fig-1}
 \end{figure}

 \begin{figure}[ht]
 %\vspace{0.3cm}
 \centering
 {\centering \resizebox*{12cm}{9cm}{\rotatebox{-90}{\includegraphics{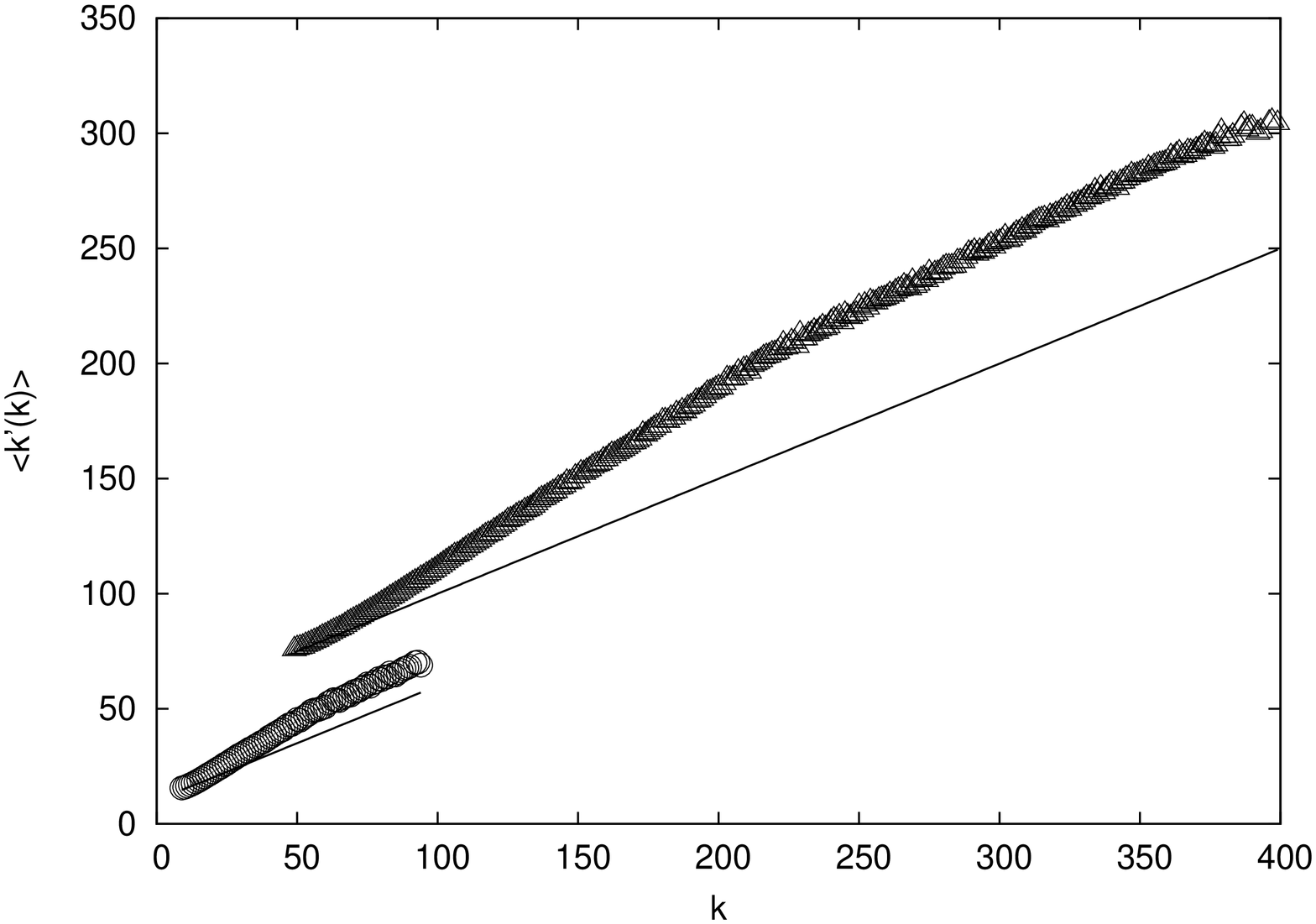}}}}
 %\vspace{0.3cm}
 \caption{Degree-degree correlations in the line graphs constructed from the growing exponential networks. The data shown are obtained for $\left\langle k\right\rangle=10$ and $50$ (circles and triangles, respectively). Lines are obtained from Eq. 3.}
 \label{fig-1}
 \end{figure}

 \begin{figure}[ht]
 %\vspace{0.3cm}
 \centering
 {\centering \resizebox*{12cm}{9cm}{\rotatebox{-90}{\includegraphics{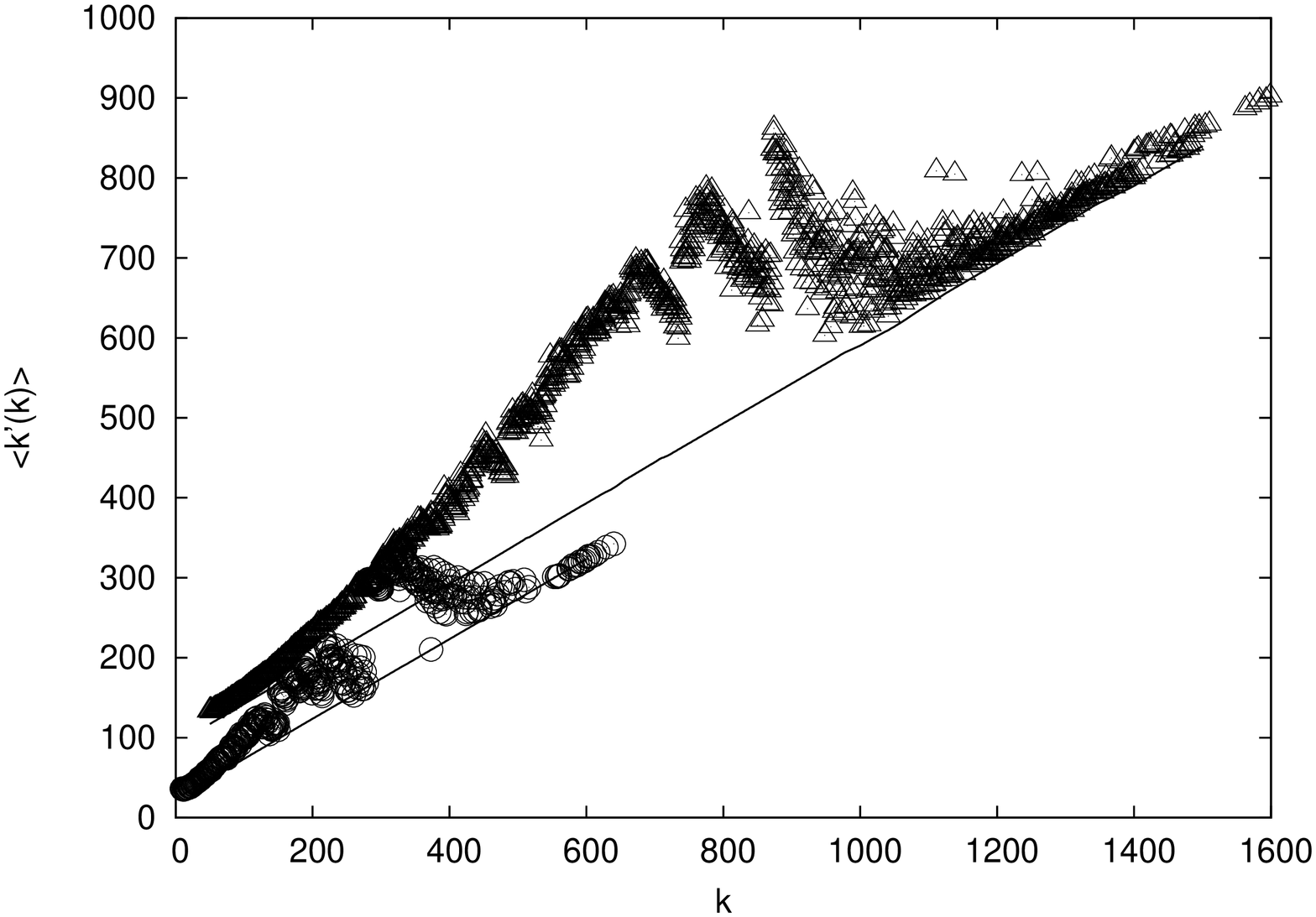}}}}
 %\vspace{0.3cm}
 \caption{Degree-degree correlations in the line graphs constructed from the growing Barab\'asi-Albert networks. The data shown are obtained for $\left\langle k\right\rangle=10$ and $50$ (circles and triangles, respectively). Lines are obtained from Eq. 1.}
 \label{fig-1}
 \end{figure}

For the Erd\"os-R\'enyi networks $P(k)$ is Poissonian; let us denote $<k>=\lambda$. We get

\begin{equation}
\left\langle k'(k)\right\rangle=\lambda+1+k-\frac{2^{k-1}(2+k)-1-k}{2^k-1}
\end{equation}
what is close to $\lambda+k/2$ for large $k$. \\

For the exponential networks with the minimal degree $M$ the degree distribution is $P(k)\propto c^k$, what gives $<k>=2M$, $c=M/(1+M)$ and

\begin{equation}
<k'(k)>=\frac{2k+5<k>-2}{4}
\end{equation}
In this case the sums in Eq. 1 start from $k_i=M$, $i=1,2,3$. After eliminating the sum over $k_2$, the sum over $k_3$ ends at $k_3=k-M+2$.\\

For the scale-free networks $P(k)\propto k^{-3}$ and the obtained series does not converge. For finite networks we can use Eq. 1 with an upper 
cut-off of $k_1$, determined by the system size \cite{bopasv}. The obtained plot is practically the same for the cut-off between $10^3$ and $10^4$. 
The limits of summations are the same as for the exponential networks.  \\

\section{Numerical calculations for the line graphs}

The line graphs are constructed from the initial networks as follows. In the connectivity matrix of the initial network, the number of units
above the main diagonal are substituded by their consecutive numbers. The maximal number is equal to the number of nodes in the line graph.
In the connectivity matrix of the line graph, two nodes $i$ and $j$ are linked if the numbers $i$ and $j$ are in the same row or the same column
in the renumbered connectivity matrix of the initial network. The same algorithm of construction of the line graphs was applied in \cite{my}.\\

The size of the initial network is equal $10^4$. The calculations are performed for the line graphs of size dependent on the size, type and connectivity of the initial network. Then, the line graphs constructed from the Erd\"os-R\'enyi networks of the mean degree $\left\langle k\right\rangle=5,10,20$ and $50$
are of size of $25,50,100$ and $250$ thousands, respectively. For the initial exponential and the Barab\'asi-Albert networks of degree $\left\langle k\right\rangle=4,10,20$ and $50$ the sizes of the line graphs are respectively $20,50,100$ and $250$ thousands. The degree distribution of the obtained line graphs was described
in details in \cite{my}; briefly, the line graphs retain the degree distributions of the initial networks.\\

The degree-degree correlations in the line graphs, obtained numerically, are shown in Figs. 4, 5 and 6 for the initial networks of three kinds:
the Erd\"os-R\'enyi networks, the exponential networks and the Barab\'asi-Albert networks, respectively. In the same graphs the theoretical curves are shown, derived from Eq. 1 with an assumption, that there is no degree-degree correlations in the initial networks. However, as we see in Figs. 1, 2 and 3, this assumption is perfectly true only in the case of the Erd\"os-R\'enyi networks. Then it is not surprising, that the numerical results 
on the degree-degree correlations agree perfectly with theory only for this kind of networks (Fig. 4). As the exponential networks show assortativity
(Fig. 2), the degree-degree correlations in the line graphs formed from the exponential networks differ from the theoretical data (Fig. 5).
Finally, the noisy character of the correlations in the initial scale-free networks, observed in Fig. 3, has some counterpart in Fig. 6. Moreover, in the latter case the numerical curves show some systematic deviation from theory till some value of the degree $k$. One of possible exploanations of the
observed deviations could be the influence of hubs. We checked that this part of data differ from one generated graph to another.

\section{Conclusions}

Our numerical results on the $\left\langle k'(k)\right\rangle$ for the original networks indicate that the degree-degree correlations are remarkable for the exponential 
networks, but they are negligible for the Erd\"os-R\'enyi networks and the Barab\'asi-Albert scale-free networks as long as the mean degree
is large enough. These results coincide with the former calculations of the clustering coefficient $C$ \cite{my}, where the largest difference between
theoretical and numerical results were found for the exponential networks. These results agree also with analytical calculations of other authors
\cite{new2,new,qi}.\\

The degree-degree correlations in the exponential networks allow to interpret also the results on the $\left\langle k'(k)\right\rangle$ dependence in the line graphs.
As before, the theoretical calculations are performed with the assumption that the correlations are absent in the initial networks. 
We know that this assumption is not true for the exponential networks. As a result, the theoretical curves $\left\langle k'(k)\right\rangle$ for the line graphs 
formed from the exponential networks differ from the same curves obtained from the numerical simulations. On the contrary, the accordance is 
quite good for the Erd\"os-R\'enyi networks, where the degree correlations are absent. For the scale-free networks of finite size, theory 
gives a linear plot $\left\langle k'(k)\right\rangle$. The simulation for these networks gives a broad distribution of points, and therefore the accordance is only qualitative. Summarizing, all the investigated line graphs are assortative. These degree-degree correlations can be understood as a consequence of the fact
that the neighboring nodes in the line graphs are formed from links sharing a common node in the initial graph. The degree of this common node
contributes to the degree of both neighboring nodes in the line graph. \\

% \section*{Acknowledgements} The calculations were performed in the ACK Cyfronet, Cracow, grants No. MNiSW/SGI3700 /AGH /030/ 2007
% and MNiSW/SGI3700/AGH /031/ 2007. This work was partially supported from the AGH UST project No. .......

\end{document}